\documentclass[conference]{IEEEtran}
\usepackage{balance}
\usepackage{enumitem}
\usepackage{mathtools}
\usepackage{graphicx}
\usepackage{epstopdf}
\usepackage{amsmath,bm}
\usepackage{float}
\usepackage{amssymb}
\usepackage{cite}
\usepackage{multirow}
\usepackage{booktabs}
\usepackage{subfigure}
\usepackage{array}
\usepackage{lipsum}                     
\usepackage{xargs}    
\usepackage[disable]{todonotes}
\usepackage{balance}
\usepackage{color,soul}
\usepackage{hyperref}
\normalsize

\ifCLASSINFOpdf
\else
\fi

\begin{document}
%
\title{Cooperative Object Detection and Parameter Estimation Using Visible Light Communications}
%
%
\author{\IEEEauthorblockN{Hamid Hosseinianfar\IEEEauthorrefmark{1}~\href{https://orcid.org/0000-0001-7998-138X}{\includegraphics[scale=0.016]{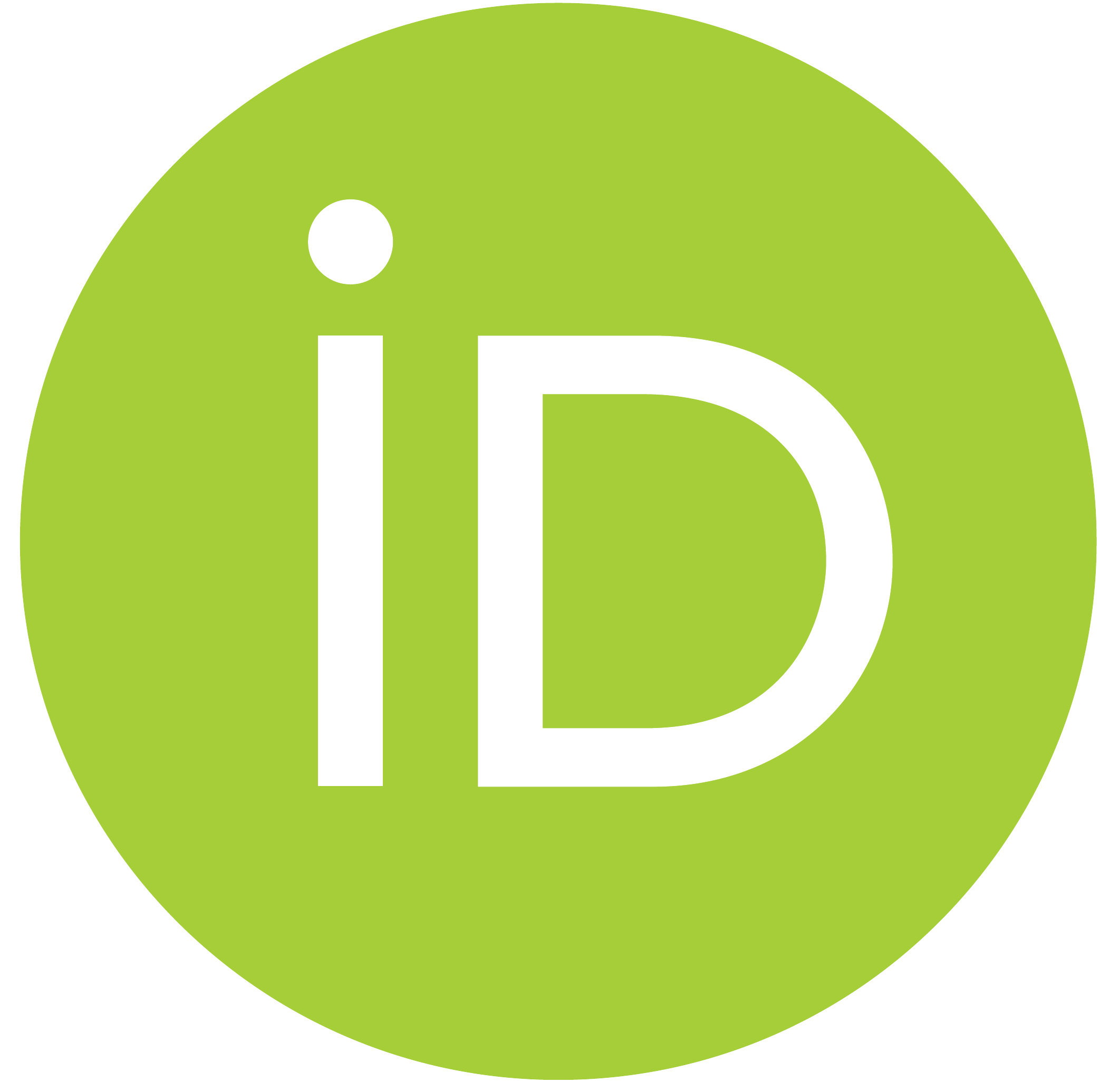}}, and Maite Brandt-Pearce\IEEEauthorrefmark{3}~\href{https://orcid.org/0000-0002-2566-8280}{\includegraphics[scale=0.016]{Figures/ORCID-iD.pdf}}\\}
	\IEEEauthorblockA{Charles L. Brown Department of Electrical and Computer Engineering, \\ University of Virginia, Charlottesville, VA 22904.\\}
		Email: \IEEEauthorrefmark{1}hh9af@virginia.edu,
		\IEEEauthorrefmark{3}mb-p@virginia.edu}

\maketitle

\begin{abstract}
Visible light communication (VLC) systems are promising candidates for future indoor access and peer-to-peer networks. The performance of these systems, however, is vulnerable to the line of sight (LOS) link blockage due to objects inside the room. In this paper, we develop a probabilistic object detection method that takes advantage of the blockage status of the LOS  links between the user devices and transceivers on the ceiling to locate those objects. The target objects are modeled as cylinders with random radii. The location and size of an object can be estimated by using a quadratic programming approach. Simulation results show that the root-mean-squared error can be less than $1$ cm and $8$ cm for estimating the center and the radius of the object, respectively.
 
\end{abstract}

\section{Introduction}
\label{sec:introduction}
\todo[inline]{Human mobility model and dataset\cite{liu2018temporal}}
\todo[inline]{Dirichlet distribution for mobility}
\todo[inline]{CCNC_R1:The work relies extensively in complex equations, which are not always trivial to follow, and should therefore be better explained throughout the paper.
Additionally, the authors consider in their system some constraints that may not be realistic, and which effects could alter the results for the proposed solution. As an example, I think understanding the signal to noise relation to the designed solution could further enhance the robustness of the solution.
Finally, it would be interesting to see a comparison of the author's results to those obtained by similar proposals (such as from the references [15] and [16]).}
\todo[inline]{* real-time algorithm of channel learning with low computational complexity}
\todo[inline]{Not very clear how their proposed work advances the state of the art. For indoor localization, only their own work is reviewed. Regarding the results, no benchmark is shown.}
\todo[inline]{They claimed for their tested area the accuracy in the magnitude of cm can be achieved. However, it is not clear what are for the requirement of localization accuracy for indoor applications.}
\IEEEPARstart{T}{he} ubiquitous use of light-emitting diodes (LED) for lighting purposes has motivated researchers to consider visible light communications (VLC) technology in future developments of indoor wireless access networks because of its potential to establish high throughput, secure, and low latency data transmission \cite{noshad2013can,lian2018optical,6876267}. Due to the micron-scale wavelength of the light signals, VLC systems facilitate the emergence of other promising technologies, such as indoor localization \cite{zhang2013comparison} and occupancy detection \cite{7917871}. However, VLC systems are mostly dependent on the line-of-sight (LOS) because their signals cannot transmit through or diffract around ordinary objects such as furniture and humans. This feature makes these systems susceptible to shadowing and blockage due to opaque objects in the room, where this effect can lead to nearly $70$~dB of extra signal attenuation \cite{xiang2014human}. This paper addresses this problem by using the VLC signals themselves to locate objects that cast a shadow in the indoor space, so as to improve the VLC performance.

From a communications perspective,  most research studies address the blockage problem by introducing robust multiplexing techniques. For example, Guzman et al. \cite{7389772, 8169113} introduce robust modulation techniques to decrease the blockage probability while still guaranteeing a tolerable level of co-channel interference for an optical atto-cell network. Adaptive joint modulation is another multiplexing technique to maximize the throughput of the network, considering the blockage as well as illumination constraints \cite{PAJO}. In both of these techniques, knowing the location and dimensions of the obstacles inside the room enables the network to run a more robust and efficient resource allocation algorithm. This knowledge also paves the way for the design of more efficient cellular structures and handover algorithms.

Using communication signals for indoor sensing and localization has recently become a hot research topic due to its applications in smart buildings. Studies on RF systems have proposed methods to detect the presence of humans and determine their location inside the room by using discriminant features of the channel state information \cite{8645524}. In \cite{8647875}, a VLC-based passive indoor localization approach is developed in which the location of the objects can be estimated using features of the impulse response between the photodetectors and LEDs on the ceiling. Another VLC-based object detection method that estimates the objects' parameters using the LOS is introduced in \cite{jarchlo2018li}. The focus of this work is on enhancing the resolution of the reconstructed image of the objects.

\todo[inline, color=green]{1. Your motivation is to develop a real-time algorithm, however, you didn't emphasis the real-time. In other words, the computational complexity is not discussed. Since you mentioned to focus on the simplification of the algorithm, but what is the complexity.  }

In this paper, we develop an algorithm to detect obstacles inside a room with a focus on realizable and straightforward solutions.  For this purpose, we consider the LOS blockage between the user equipment (UE) transmitter and a network of uplink receivers on the ceiling as the observation dataset. By using this dataset, the object detection algorithm can estimate the location of the center and the radius of an obstacle using a quadratic programming approach. In this case, the object with the most probable shadowing effect can be detected by the VLC signals themselves. This technique will play a fundamental role in future indoor Internet of things (IoT) networks by enabling straightforward awareness of the situation and the surrounding area.

The rest of the paper is organized as follows. In Section~\ref{Sec:System_Description}, the system model is discussed. The cooperative object detection algorithm is derived in Section~\ref{Sec:Object_detection}. Numerical results are presented and discussed in Section~\ref{Sec:Numerical}. Finally, the paper is concluded in Section~\ref{Sec:Conclusion}.
\section{System Description}
\label{Sec:System_Description}
\subsection{System Overview}
\begin{figure}[t]
	\centering
\begin{tabular}{c}
\includegraphics[width=2.95in]{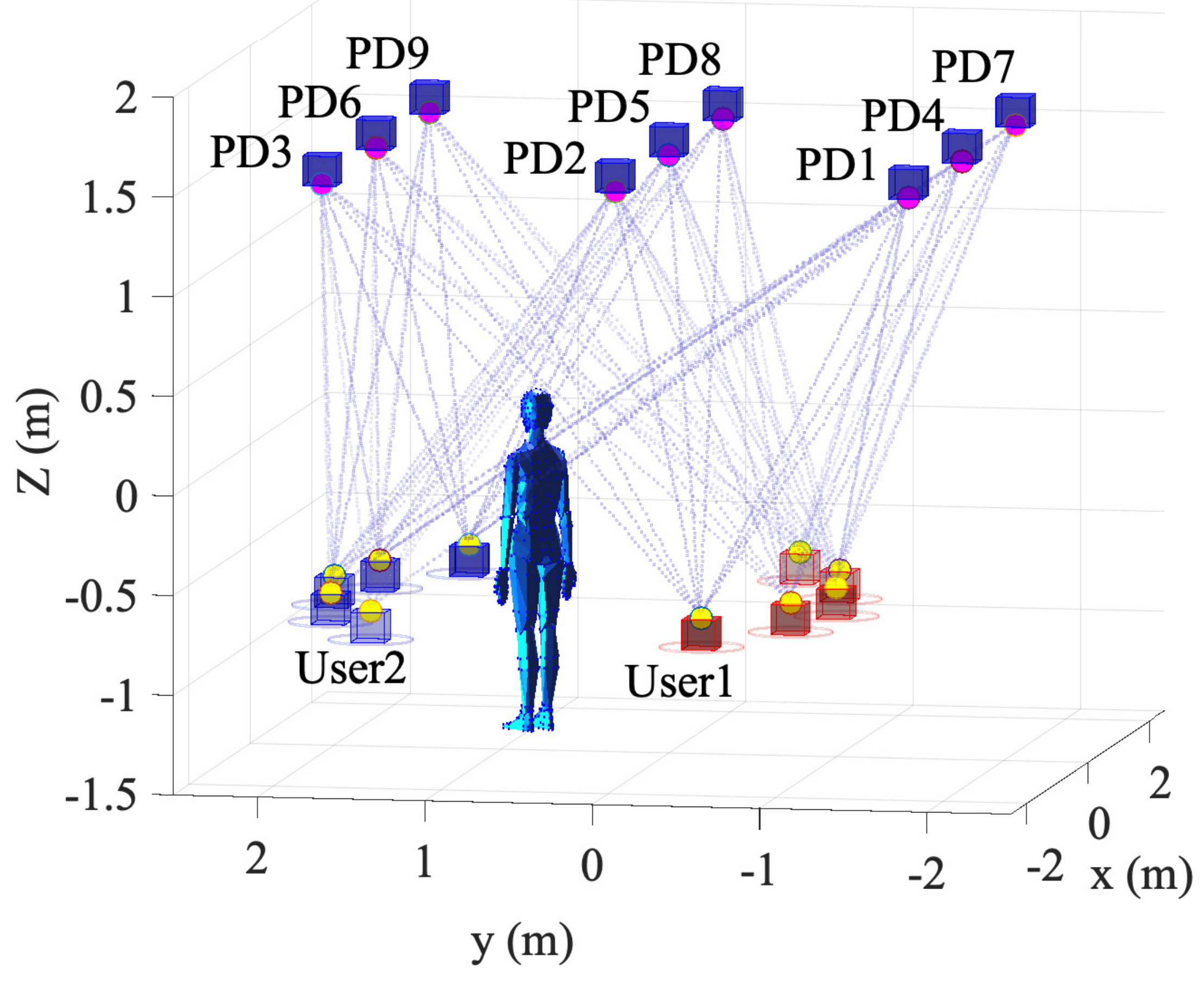}\\ (a)
\\
\includegraphics[width=2.9in]{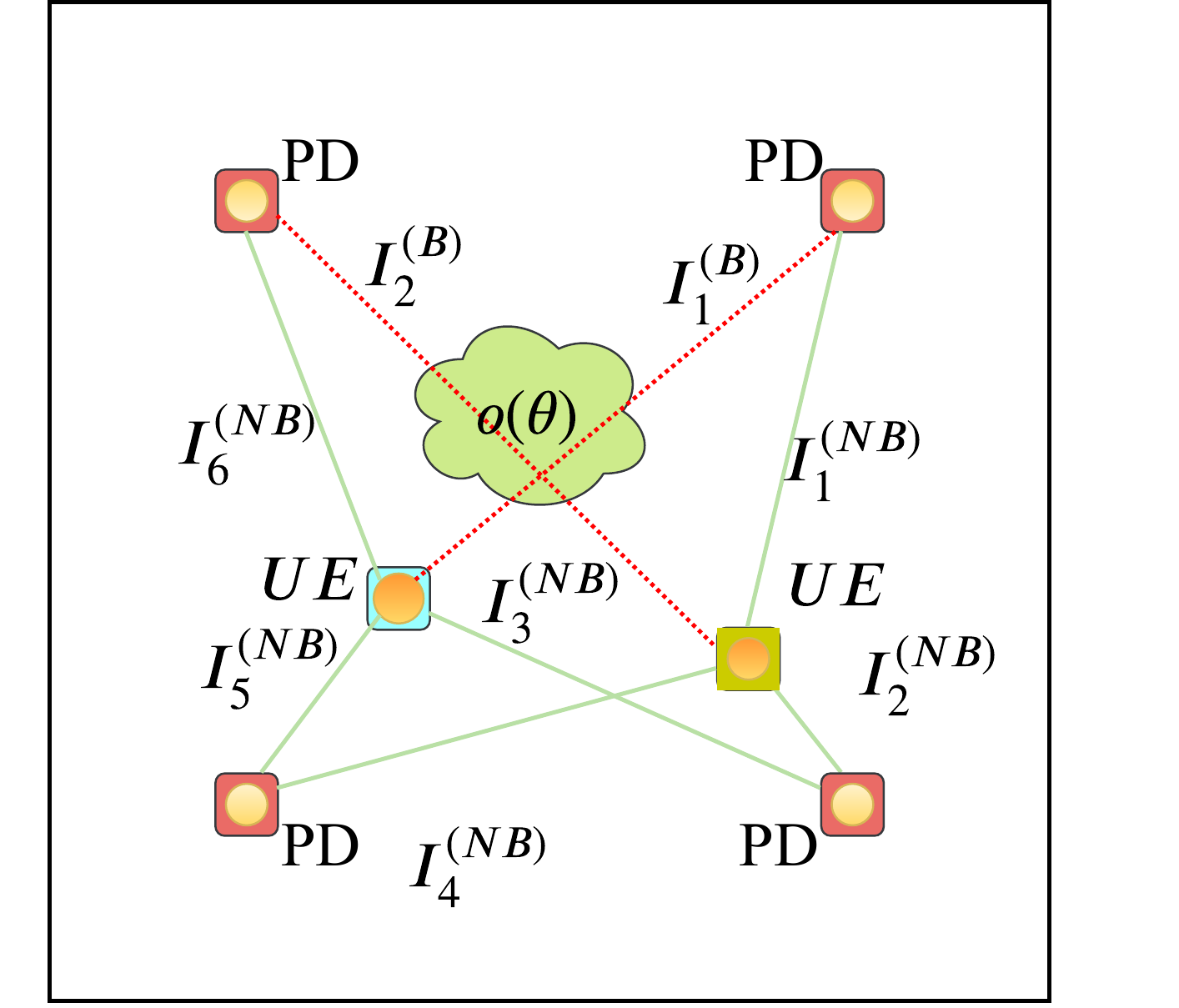} \\
(b)
\end{tabular}
	\caption{(a) Multiple moving users and  (b) bird's-eye view illustrating the object detection algorithm that uses LOS blockage information}
	\label{fig:Scenarios} 
\end{figure}

\todo[inline]{Hamid: Explain that use use high snr and coded signal to differentiate the UE MEasurementns.}
In this work, we consider a typical VLC system with a  grid of photodetectors (PDs) on the ceiling. The UE can be any device equipped with an optical transceiver, such as a cell phone, a wearable gadget, a laptop dongle, an autonomous agent, or an IoT device.
\todo[inline, color=yellow]{Hamid: problem with flow of story }
The algorithm collects blockage information from the signal status of users while they are moving inside the room, as illustrated in Fig.~\ref{fig:Scenarios}-(a).  The proposed algorithm can detect large objects inside the room based on the blockage of the LOS signal between each node on the ceiling and the UEs.

Fig.~\ref{fig:Scenarios}-(b) demonstrates a top-view abstraction of a typical scenario of VLC signal blockage. Our observations in the proposed algorithm are all the blockage indicators,
\begin{equation}
    \mathbf{I}=\left \{ I^{(B)}_1, \cdots, I^{(B)}_N ,I^{(NB)}_1, \cdots, I^{(NB)}_M\right \}
    \label{Eq:Obs_Set}
\end{equation}
where $N$ and $M$ are the total numbers of blocked (B) links and non-block (NB) links. In this work, we assume that the signal to noise ratio is sufficiently large that the observations $I_i^{(\cdot)}$  can be obtained with zero error. The algorithm employs orthogonal codes for each UE transmission to identify all the B-link and NB-link indicators. 
\todo[inline, color=yellow]{Hamid: No connection.}

We further assume that, as the multiple UEs move inside the room, a substantial portion of the indoor space can be covered. By using the link-level information, the proposed algorithm can detect objects that degrade the performance of the communication system for users located near the UEs themselves, i.e., only  objects relevant to real-time communication needs are detected.  
\todo[inline, color=green]{Refer to Fig. \ref{fig:Scenarios}-b}
Without loss of generality, we assume that the VLC network knows the  UEs' locations. This assumption is reasonable considering recent VLC indoor localization techniques that have been shown to provide centimeter-level positioning accuracy \cite{zhang2013comparison,hosseinianfar2017positioning}. Based on the location of the PDs on the ceiling and the UEs, the geometric parameters of LOS links can easily be determined. 

\subsection{Object Model}
\section{Object Model}
\todo[inline]{Globecomm: In section IV before Table I, you should discuss why and when practical objects are well represented by your cylinder approximation.}
A typical obstacle inside a room, such as a human or a column, can be modeled as a cylinder with a random radius with Gaussian distribution $\mathcal{N}\left ( \mu_r, \sigma_r \right )$, where $\mu_r$ and $\sigma_r^2$ are the mean and variance of the object radius, respectively, which are assumed known to the detection algorithm. This model is based on the fact that the Gaussian distribution is the best fit for a random parameter distribution with known mean and variance when there is no more specific information. In our problem if the object is a person, we do not know their age, gender, race, and height, etc. The normal distribution is also validated as a model of waist circumference for healthy men and women in the United States \cite{flegal2007waist}.  The location of the center of object is a second estimation parameter, denoted as $\boldsymbol{\theta}=({\theta}_{x}, {\theta}_{y})$.

\begin{figure}
    \centering
		\includegraphics[width=2.5in]{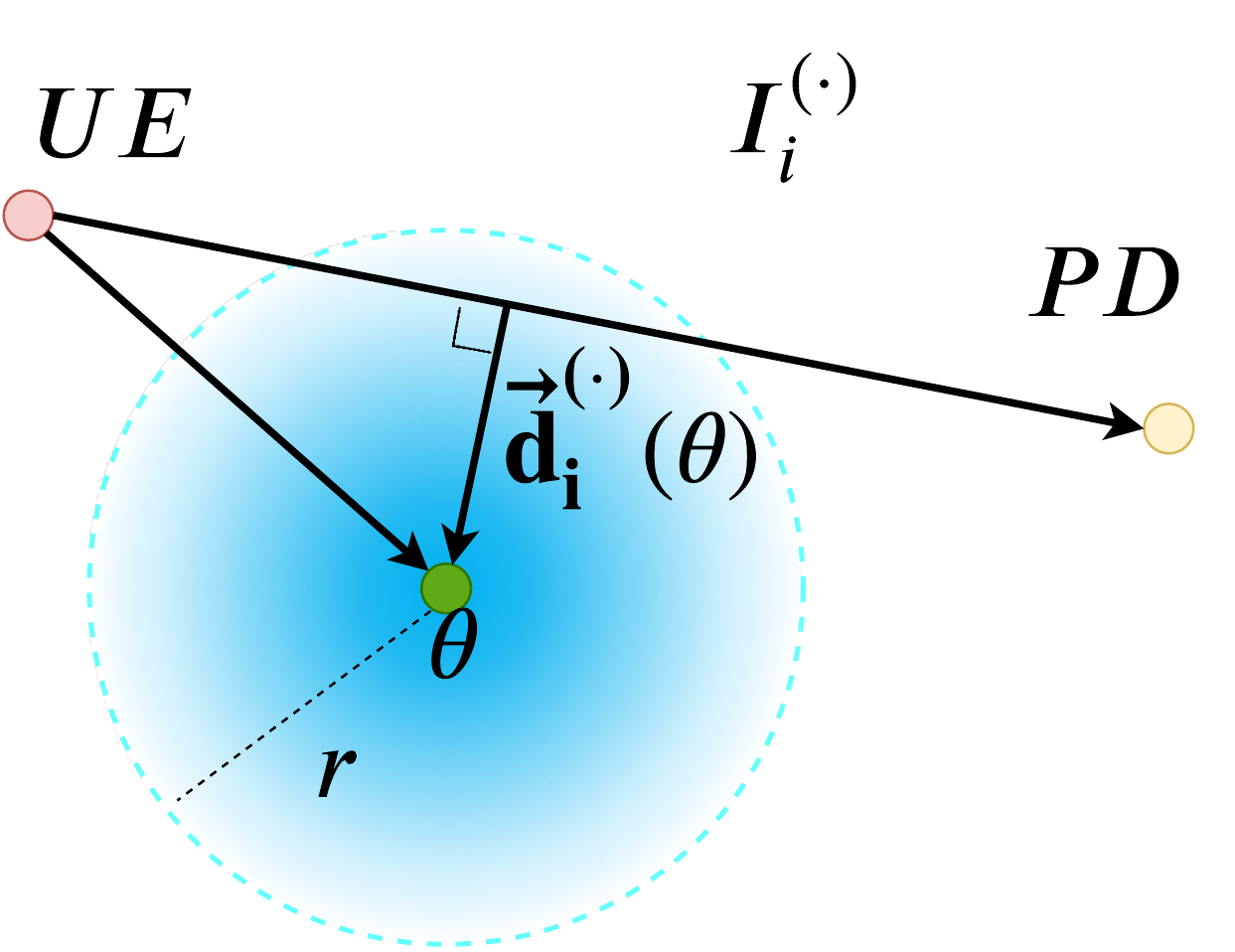} 
		\caption{Distance of an object from a LOS link}
		\label{Fig:OL_Distance_Circle}
\end{figure}

Fig. \ref{Fig:OL_Distance_Circle} illustrates the distance of an object with a random radius $r$  from a LOS link. The blockage indicator $I^{(B)}_i$  measured by our algorithm shows that there is an object close to the line that defines the LOS. However, this distance is not deterministic due to the random size and shape of the object. Even if the algorithm knows the radius and shape of the object, it only knows that the distance between the line and the center of the object, $\boldsymbol{\theta}$, is less than its radius, but there is no clue about the exact distance from the center of the object. Considering this fact, the posterior probability of $\boldsymbol{\theta}$ can be written as
\begin{align}
 &p(\boldsymbol{\theta}|I^{(B)}_i)=p\left(\left |\mathbf{\vec{d}^{(B)}_{i}}(\boldsymbol{\theta}) \right |\leq r\right)=\nonumber \\&\int_{\left |\mathbf{\vec{d}^{(B)}_{i}}(\boldsymbol{\theta}) \right |}^{+\infty }\frac{1}{\sqrt{2\pi\sigma_r^2}}e^{-\frac{(t-\mu_r)}{2\sigma_r^2}}dt \propto Q\left (\frac{\left |\mathbf{\vec{d}^{(B)}_{i}}(\boldsymbol{\theta}) \right |-\mu_r}{\sigma_r}  \right )\nonumber \\
& p(\boldsymbol{\theta}|I^{(NB)}_i)=p\left(\left |\mathbf{\vec{d}^{(NB)}_{i}}(\boldsymbol{\theta})  \right |>  r\right)=\nonumber \\
&\int_{0}^{\left |\mathbf{\vec{d}^{(NB)}_{i}}(\boldsymbol{\theta})  \right |}\frac{1}{\sqrt{2\pi\sigma_r^2}}e^{-\frac{(t-\mu_r)}{2\sigma_r^2}}dt\propto 1- Q\left (\frac{\left |\mathbf{\vec{d}^{(NB)}_{i}}(\boldsymbol{\theta})  \right |-\mu_r}{\sigma_r}  \right )
\end{align}
where $\mathbf{\vec{d}^{(\cdot)}_{i}}(\boldsymbol{\theta}) $ is the vector of 2-D Euclidean distances from $\boldsymbol{\theta}$ to the $i$th link, and the superscripts $(B)$ and $(NB)$ indicate the parameters corresponding to blocked-links and nonblocked-links. Considering the parameters of the $i$th link LOS, $\mathbf{\vec{n}}^{(\cdot)}_{i}$,  and the distance between the link and the origin of the room's coordinate system, $\beta^{(\cdot)}_{i}$,  we can calculate this distance as   
\begin{equation}
	\mathbf{\vec{d}^{(\cdot)}_{i}}(\boldsymbol{\theta})=(\mathbf{\vec{n}}^{(\cdot)}_{i}\cdot \boldsymbol{\theta}-\beta^{(\cdot)}_{i}) {\mathbf{\vec{n}}^{(\cdot)}_{i}}
\end{equation}


\section{Object Detection Algorithm}
\label{Sec:Object_detection}
\todo[inline]{Hamid: Explain here that this system is scalable to multiple users}
In this section, we discuss the optimum solution for estimating the location and size of the objects. We further propose a simple sub-optimal estimation algorithm that has significantly lower complexity. In developing these estimation algorithms, we consider the presence of a single object inside the room. In addition, the algorithms estimate the object location in a 2-D plane; the algorithms are directly scalable to 3-D object detection and multiple objects scenarios, the details of which are left for future work.
\subsection{Maximum Likelihood Object Detection}
\label{subSec:ML}
Given the set of measurements, $\mathbf{I}$, defined in (\ref{Eq:Obs_Set}), the optimum estimation for $\boldsymbol{\theta}$, the coordinates of the center of the object on the $xy$-plane, can be obtained based on the maximum likelihood criteria as
\begin{align}
\hat{\boldsymbol{\theta}}= \underset{\boldsymbol{\theta}  \in \boldsymbol{\Theta}}{\arg  \max}(p(\boldsymbol{\theta}|\mathbf{I}))
\label{Eq:Likelihood}
\end{align}
Assuming the  LOS link parameters $(\mathbf{\vec{n}}_{(i)},\beta_{i})$ are independent, the likelihood of all measurement can be written as
\begin{align}
p(\boldsymbol{\theta}|\mathbf{I})= &\prod_{i=1}^{N}\prod_{j=1}^{M}Q\left (\frac{\left |\mathbf{\vec{d}}^B_{i}(\boldsymbol{\theta}) \right |-\mu_r}{\sigma_r}  \right )\nonumber \\
& \times\left ( 1-Q\left (\frac{\left |\mathbf{\vec{d}}^{NB}_{j}(\boldsymbol{\theta}) \right |-\mu_r}{\sigma_r}  \right ) \right )
\end{align}
Obtaining a closed-form solution for this problem is computationally complex.

\subsection{MMSE Object Detection}
\label{subSec:MMSE}

In most applications, having a rough estimate of $\boldsymbol{\theta}$ is sufficient. In this work, we concentrate on developing an efficient single object detection algorithm that is scalable to multi-object detection. In this section, we simplify the object location problem by using a constrained minimum mean square error (MMSE) approach based on realistic assumptions.

For enough number of measurements, we anticipate that  $\mathbf{E}\left [\mathbf{\vec{d}}^B_{i}(\boldsymbol{\theta}^*)\right ]=0$, in which $\boldsymbol{\theta}^{*}$ is the optimum $\boldsymbol{\theta}$. With this assumption, the problem can be simplified to an MMSE problem. In addition, the NB-link measurements can be deployed as constraints for the MMSE objective function. The optimization algorithm can then be written as 
\begin{align}
\hat{\boldsymbol{\theta}}= &\underset{\boldsymbol{\theta}  \in \boldsymbol{\Theta}}{\arg  \min}\sum_{i=1}^{N} \mathbf{\vec{d}}^B_{i}(\boldsymbol{\theta})\cdot{\mathbf{\vec{d}}^B_{i}(\boldsymbol{\theta})}^T \nonumber \\
& s.t.: \left | d^{NB}_{j}(\boldsymbol{\theta}) \right |>D^{(NB)}_{min} ,\quad  j\in \left \{ 1,\cdots,M \right \}
\label{Eq:MMSE}
\end{align}
where the algorithm parameter $D^{(NB)}_{min}=\mu_r+\alpha \sigma_r$ is the guard distance of $\hat{\boldsymbol{\theta}}$ from the NB-links. The initial value of  $\alpha=3$ is set to guarantee that $D^{(NB)}_{min}$ is larger than all the possible object's radii. The algorithm might have a null feasible region due to a large given $D^{(NB)}_{min}$. In this case, $\alpha$ can be reduced iteratively until the NB-links' constraints create a non-null feasible region. 

This constrained optimization problem is a quadratic programming problem that can be solved based on the Karush-Kuhn-Tucker (KKT) criteria \cite{boyd2004convex}. 
In order to find the global optimum solution to this problem, the optimum point $\boldsymbol{\theta}^{*}$, we define the Lagrangian function as follows:
\begin{align}
\mathcal{L} (\boldsymbol{\theta},\boldsymbol{\lambda},\boldsymbol{\gamma}):=&\sum_{i=1}^{N}(\mathbf{n}^{(B)}_i\cdot\boldsymbol{\theta}_k-\beta^{(B)}_i)^2\nonumber\\ &+\sum_{j=1}^{M} \lambda_{j}\left [D^{(NB)}_{min} -\left (\mathbf{n}^{(NB)}_j\cdot\boldsymbol{\theta}-\beta^{(NB)}_j  \right )  \right ]\nonumber\\ 
&+\sum_{j=1}^{M}\gamma_{j}\left [D^{(NB)}_{min} +\left (\mathbf{n}^{(NB)}_j\cdot\boldsymbol{\theta}-\beta^{(NB)}_j  \right )  \right ]
\end{align}
where $\lambda_{j}$ and $\gamma_{j}$ are the Lagrangian multipliers. $\boldsymbol{\theta^*}$ is a local minimum of the objective function if there exists a set of Lagrangian multipliers $\boldsymbol{\lambda^{*}}$, and $\boldsymbol{\gamma^{*}}$ that satisfy the following conditions, known as KKT conditions:
\begin{align}
\intertext{\textbf{Stationarity:}}
& \nabla \mathcal{L} (\boldsymbol{\theta^*}, \boldsymbol{\lambda^{*}},\boldsymbol{\gamma^{*}})=0 \label{EQ:Stationarity}\\
\intertext{\textbf{Primal and Dual Feasibility:}}
& D^{(NB)}_{min} -\left (\mathbf{n}^{(NB)}_j\cdot\boldsymbol{\theta}^{*}-\beta^{(NB)}_j  \right )<0 \quad \lambda^{*}_{j}\geq0\nonumber\\ &D^{(NB)}_{min}+\left (\mathbf{n}^{(NB)}_j\cdot\boldsymbol{\theta}^{*}-\beta^{(NB)}_j  \right )<0 \quad \gamma^{*}_{j}\geq0 \label{EQ:Feasibility}\\
&   {\rm for} \quad j=1,\cdots,M. \nonumber\\
\intertext{\textbf{Complementary slackness:}}
&\lambda^{*}_{j}\left [D^{(NB)}_{min} -\left (\mathbf{n}^{(NB)}_j\cdot\boldsymbol{\theta}^{*}-\beta^{(NB)}_j  \right )  \right ] =0, \label{EQ:slackness}\\ 
&\gamma^{*}_{j}\left [D^{(NB)}_{min} +\left (\mathbf{n}^{(NB)}_j\cdot\boldsymbol{\theta}^{*}-\beta^{(NB)}_j  \right )  \right ] =0\nonumber \\ &  {\rm for} \quad j=1,\cdots,M \nonumber
\end{align}
where $\nabla \mathcal{L}(\boldsymbol{\theta}, \boldsymbol{\lambda},\boldsymbol{\gamma})$ stands for the gradient of the Lagrangian $\mathcal{L}(\boldsymbol{\theta}, \boldsymbol{\lambda},\boldsymbol{\gamma})$. The asterisk $*$ identifies the value of the parameter that meets all KKT conditions. The stationary KKT condition  in (\ref{EQ:Stationarity}) can be written as 
\begin{align}
&  \frac{\partial \mathcal{L} (\boldsymbol{\theta}, \boldsymbol{\lambda}, \boldsymbol{\gamma})}{\partial \boldsymbol{\theta}}=2\sum_{i=1}^{N}\mathbf{n}^{(B)}_i(\mathbf{n}^{(B)}_i\cdot\boldsymbol{\theta}-\beta^{(B)}_i)\nonumber\\  &+\sum_{j=1}^{M} \left (\gamma_{j}-\lambda_{j}  \right )\mathbf{n}^{(NB)}_j=0
\label{Eq:Stationary_Simp}
\end{align}
To find the local optimum value of $\boldsymbol{\theta}$, we have to solve a linear equation system with $2+2M$ unknown parameters that satisfy the feasibility and complementary slackness conditions of (\ref{EQ:Feasibility}) and (\ref{EQ:slackness}), respectively. However, understanding the intuition behind the KKT conditions, and also the geometrical perspective of the optimization problem, helps us reduce unnecessary complexity. Considering the quadratic objective function and linear constraints, we anticipate that $\boldsymbol{\theta^*}$ is either exactly  the global optimum of the unconstrained objective function, denoted as  $\boldsymbol{\theta_g}$, or on one of the active constraints with closest Euclidean distance from $\boldsymbol{\theta_g}$. With this perspective, we can find a $\boldsymbol{\theta^*}$ that meets the KKT conditions in three steps as follows:\\

\noindent 1) \textbf{Finding the global minimum $\boldsymbol{\theta_g}$ and corresponding active constraints:} By considering all Lagrangian multipliers to be equal to zero,  $\boldsymbol{\theta_g}$ can be obtained from (\ref{Eq:Stationary_Simp}) as
    \begin{align}
    &\boldsymbol{\theta_g}=\mathbf{A}^{-1}\cdot\mathbf{\bar{\beta}}^{(B)} \nonumber \\
    &\mathbf{A}=\sum_{i=1}^{N} {\mathbf{n}^{(B)}_i}^T\cdot\mathbf{n}^{(B)}_i, \quad \bar{\beta}^{(B)}=\sum_{i=1}^{N}\beta^{(B)}_i
    \end{align}
    The active constraints corresponding to $\boldsymbol{\theta_g}$ are the constraints that do not satisfy the feasibility condition of (\ref{EQ:Feasibility}), which indicate that $\boldsymbol{\theta_g}$ is too close to some NB-link. In this case, the corresponding Lagrangian multipliers are most probably positive values, making the Lagrangian gradient zero on the border of the feasible region, i.e., creating a push vector to keep the object far away from the corresponding NB-link.\\
    
\noindent 2) \textbf{Finding the feasible intersection points of constraints:} 
    In this step, we find the intersection points of the active constraints and check the feasibility condition  (\ref{EQ:Feasibility}) for these points. The intersection points which meet feasibility condition and the points over their corresponding constraints are candidates for local minima. In the case that none of the intersection points meet the condition mentioned above, we are faced with a null feasible region; we then have to decrease $D^{(NB)}_{min}$ and repeat this step.\\
    
\noindent    3) \textbf{Solving the simplified linear equations system of KKT conditions:} Considering the quadratic objective function and anticipated linear constraints, all candidates $\boldsymbol{\theta}^*$ are located on either the closest feasible intersection point or the constraints corresponding to $\boldsymbol{\theta}_g$. Therefore, the linear equation (\ref{Eq:Stationary_Simp}) can be simplified as
    \begin{align}
        \mathbf{A}\cdot\boldsymbol{\theta}+ 0.5\mathbf{n}^{(NB)}_j\lambda_{j}= \bar{\beta}^{(B)},
    \end{align}
    where $j$ is the index of the closest link LOS to $\boldsymbol{\theta_g}$. The auxiliary equation to solve the optimization problem is the equation of the line that also satisfies the complementary slackness condition.

%

\section{Numerical Results}
\label{Sec:Numerical}
\todo[inline]{It would be better to compare existing scheme numerically with other existing schemes developed for detecting objects using LOS link blockage information.}
\todo[inline]{the reviewer think that moving obstacles should be assumed in future.}
In this section, we analyze the performance of the proposed algorithm. We consider a typical room without any furnishing to model the LOS signals.  The PDs are located on a $L \times L$ grid on the ceiling. For the $2\times 2$  PDs case, the PDs are located 1.5 m from the walls of the room. Table~\ref{Table:I} summarizes the remaining parameters used in our simulations.
\todo[inline]{The five results figures are tersely described, in a mainly narrative manner and with little motivation. }
\todo[inline]{Why is the radius estimation more accurate than the center location's estimation? Is there a physical reason behind this?}
\todo[inline]{The ranking of the curves in Figs. 3 and 4 is not clear: why, for example, do the curves for R=0.4 and R=0.6, and for theta=(0,0) intersect in Fig. 4a? And what are the main parameters to determine a ranking of the curves? What is the best RMSE as a function of location?}
\todo[inline]{The paper completely neglects the case of a moving object, which is a very typical case in indoor environments and for which it is not clear whether the proposed system would be able to collect a sufficient number of measurements or not.}
The UE transmitters are assumed to all be at the height of $85$ cm from the floor and randomly located in the $(x,y)$ plane inside the room. Considering the minimum distance between users in a realistic scenario, i.e., the personal space needed by humans, a Poisson disk sampling method popular in many computer graphics applications is employed for generating the random locations of the transmitters \cite{dunbar2006spatial}. This minimum distance is a  function of the number of users and objects inside the room.
For both simulating the link blockage and testing the algorithm, we consider a cylinder shaped obstacle with randomly generated parameters $(\boldsymbol{\theta}, r)$, which are unknown to the detection algorithm.

\begin{table}[h]
\centering
\caption{Simulation Parameters of Channel Model.}
\label{TableI}
\begin{tabular}{|l|l|}
\hline
\textbf{Transmitter Parameters} & \textbf{Value} \\ \hline
Height                                           & 0.85 m                          \\ \hline
Lambertian Mode                                  & 1                               \\ \hline
LED transmit power                               & 10 mW                           \\ \hline
\textbf{Receiver Parameters}    & \textbf{Value} \\ \hline
Surface area of the PD                           & 1 $\textrm{mm}^2$               \\ \hline
Height                                           & 3 m                             \\ \hline
Filed of View (Half Angle)                       & $70^{\circ}$                    \\ \hline
\textbf{Room Parameters}        & \textbf{Value} \\ \hline
Room Size                                        & $5\times5\times3$ m             \\ \hline
\textbf{Algorithm Parameters}   & \textbf{Value} \\ \hline
$\mu_r$                            & $0.13$                           \\ \hline
$\sigma_r$                         & $0.03$                             \\ \hline
$D^{(NB)}_{min}$             & $0.1$                             \\ \hline
\end{tabular}
\label{Table:I}
\end{table}

\begin{figure}[t]
	\centering
	\begin{subfigure}{}
	\includegraphics[width=3in]{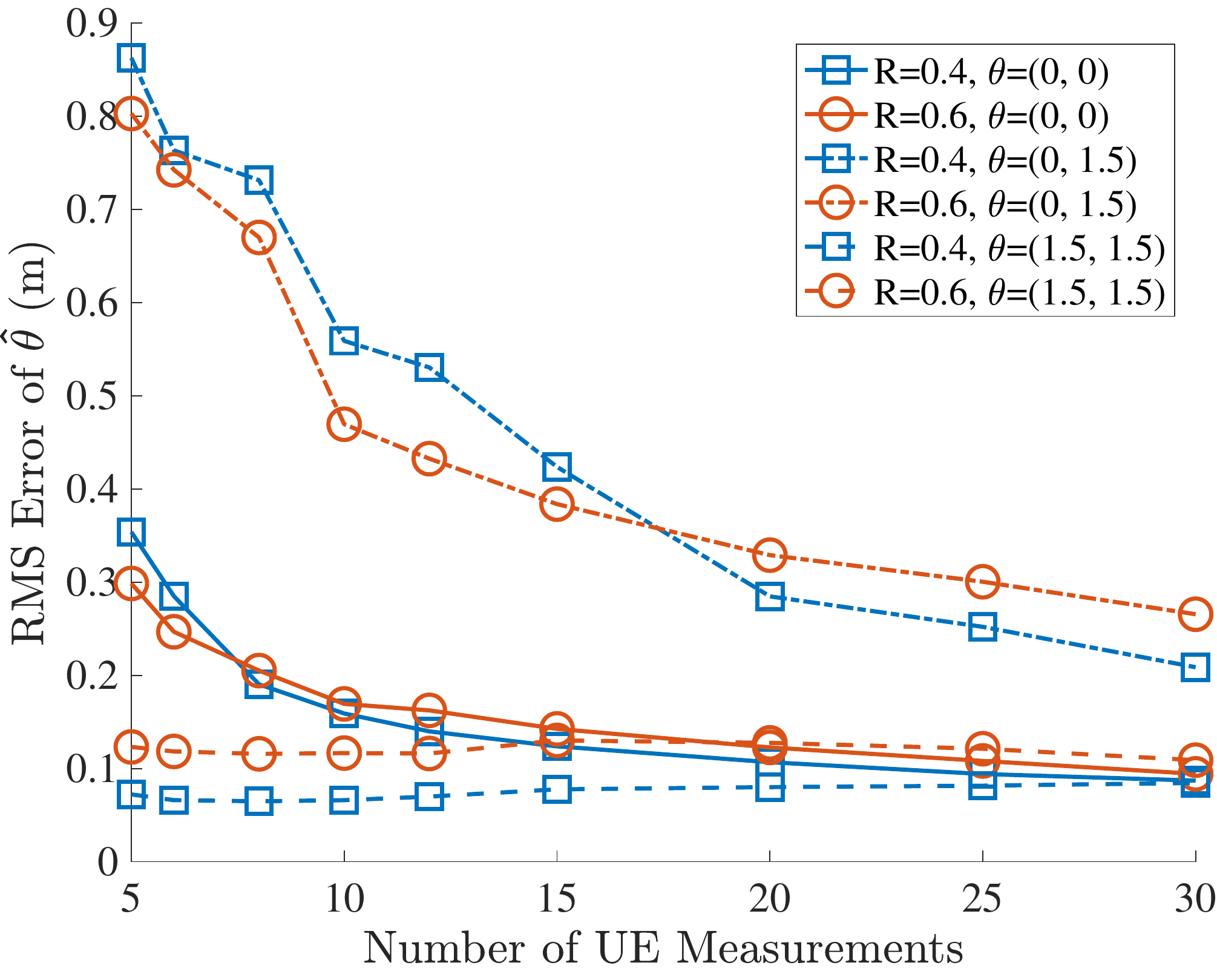}\\
	(a)
	
	\end{subfigure}
	\begin{subfigure}{}
		\includegraphics[width=3in]{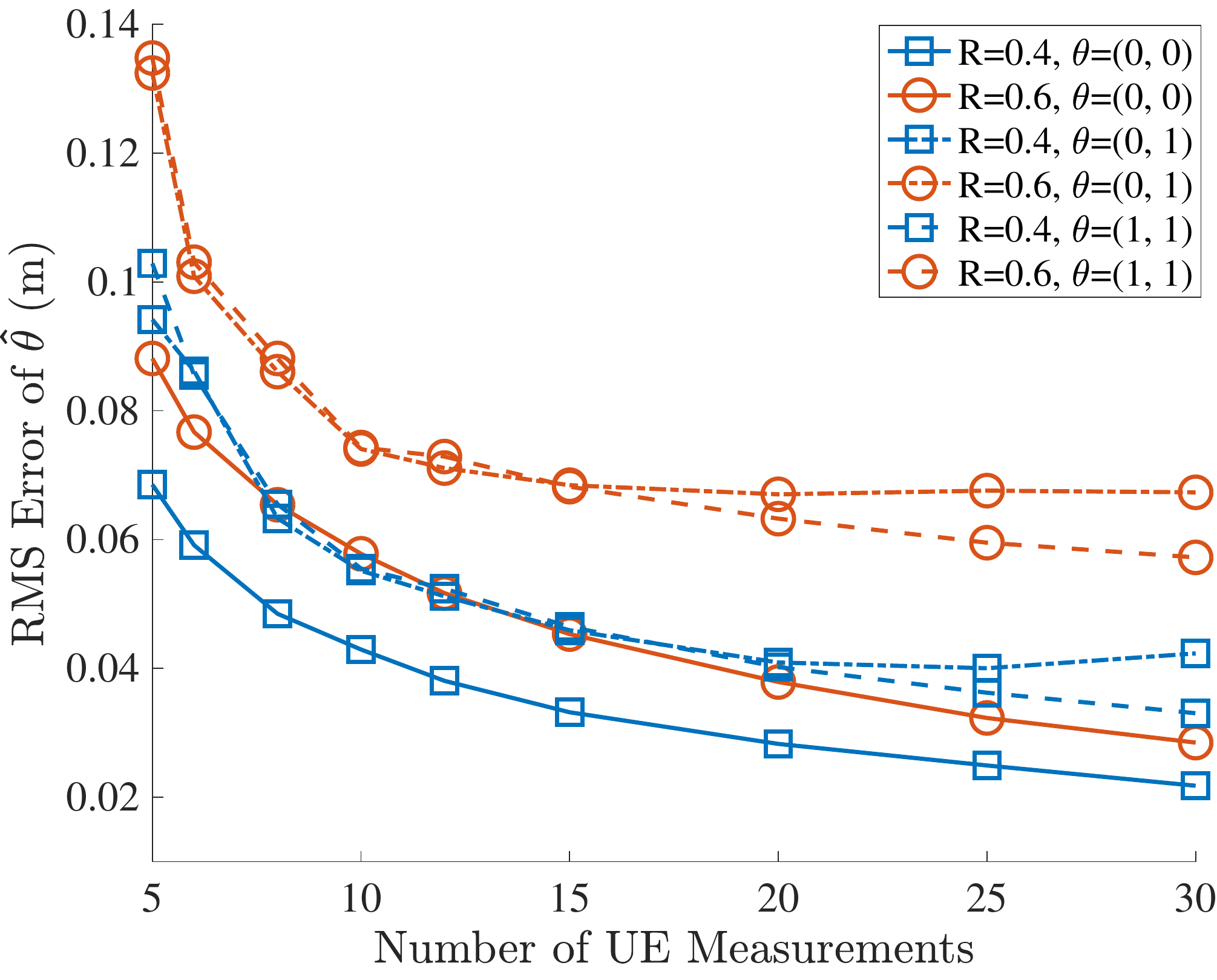}
	\\
	(b)
	\end{subfigure}
	\caption{RMS error in estimating the center of the object $\mathbf{\hat{\theta}}$: a) using a $2\times 2$  PD grid and  b) using a $5\times 5$  PD grid on the ceiling.}
	\label{fig:KKT_VS_User_Theta} 
\end{figure}
Fig. \ref{fig:KKT_VS_User_Theta} illustrates the root-mean-squared (RMS) estimation error of $\boldsymbol{\theta}$ for two different sets of PDs mounted on the ceiling versus the number of UE measurements. The algorithm can achieve an estimation accuracy of $2$ cm for the center of the object,  $\boldsymbol{\theta}$,  using a $5\times 5$ grid of PDs;  the estimation error is higher around $10$ cm for a $2\times 2$  PDs grid.

\begin{figure}[t]
	\centering
	\begin{subfigure}{}
	\includegraphics[width=3in]{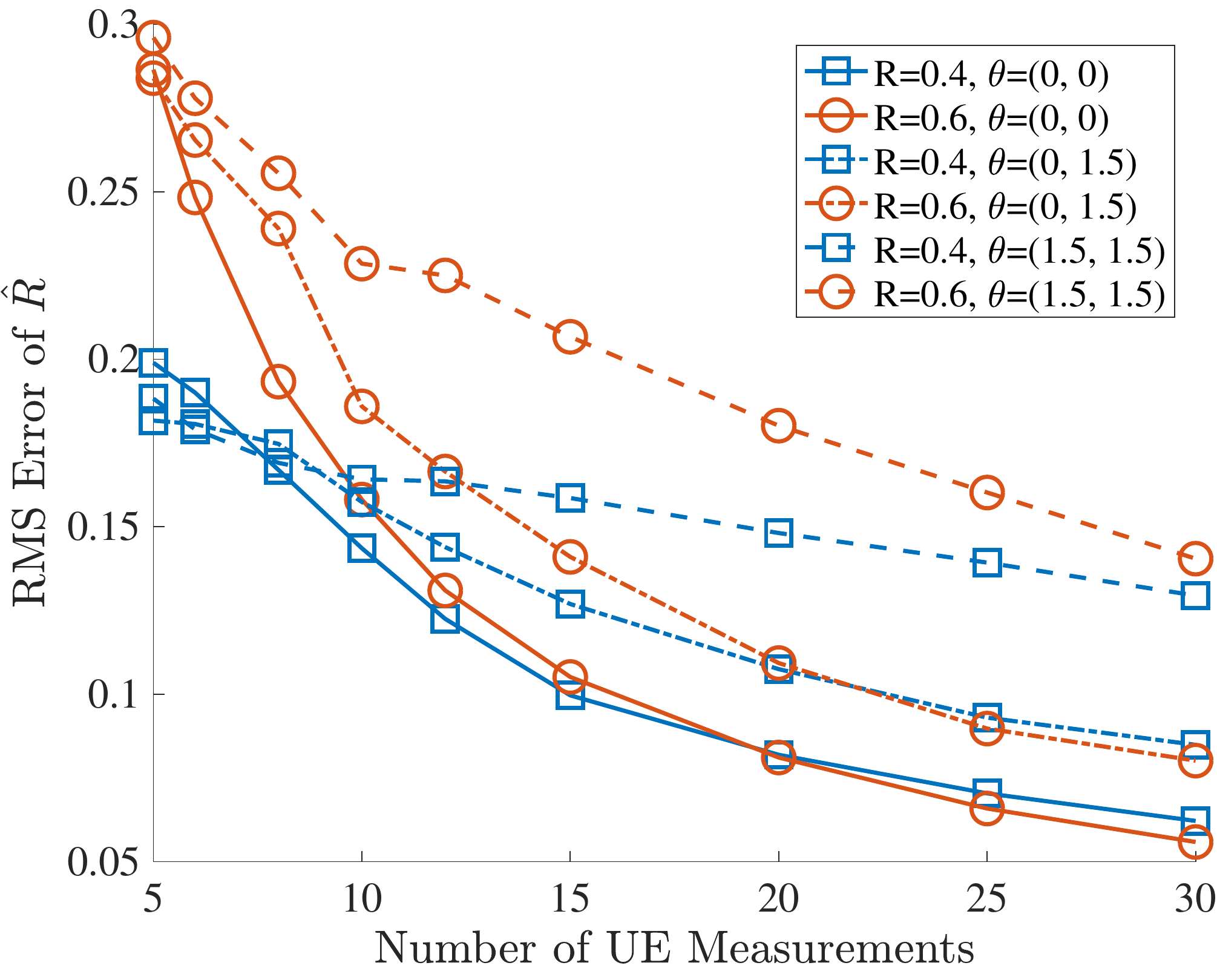}\\
	(a)
	
	\end{subfigure}
	\begin{subfigure}{}
		\includegraphics[width=3in]{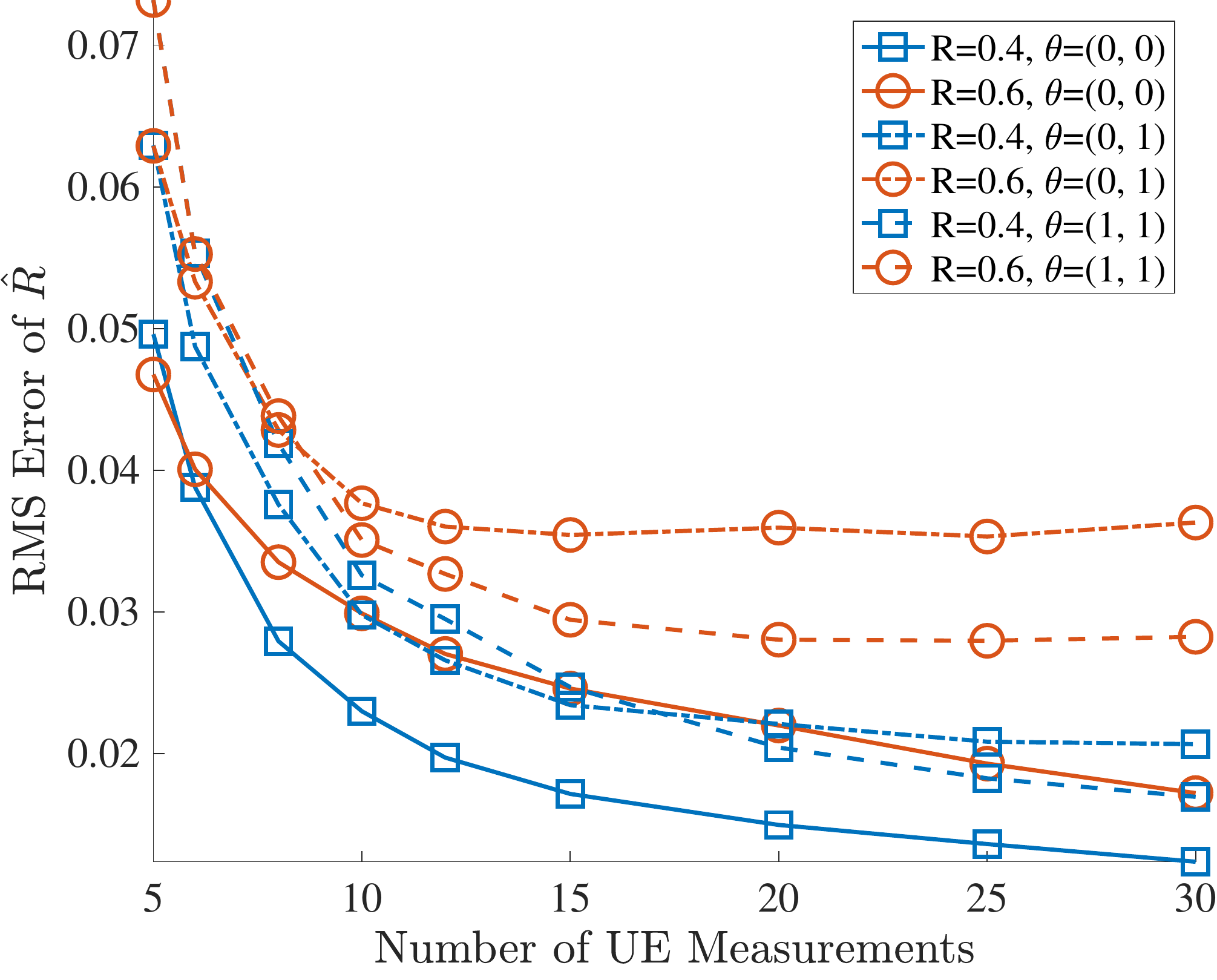}\\
		(b)
	\end{subfigure}
	\caption{RMS error in estimating the object radius $\hat{R}$: a) using a $ 2\times 2$  PD grid,  and b) using a $5\times 5$  PD grid on the ceiling.}
	\label{fig:KKT_VS_User_R} 
\end{figure}

Fig. \ref{fig:KKT_VS_User_R} shows the RMS estimation error of the radius $R$ for a different number of UE measurements, using the same sets of PDs mounted on the ceiling as for Fig.~\ref{fig:KKT_VS_User_Theta}. Similar to the estimation of the $\boldsymbol{\theta}$ parameter, the algorithm can achieve as low as $1$ cm RMS error for estimating $R$ for a $5\times 5$ PD grid and an RMS error of $5$ cm for a $2\times 2$  PDs grid scenario.

By increasing the number of UE measurements, the RMS estimation error for both the center and the radius of the object, $\boldsymbol{\theta}$, and $R$, decrease since the algorithm has more link blockage measurements.  For a sufficient number of measurements, the algorithm achieves its lowest RMS estimation error when the object is either in the center of the room or close to a PD, since the probability of blockage is higher in this regions, which leads to more measurements for the algorithm to use.

\begin{figure}[!htb]
	\centering
    \includegraphics[width=3in]{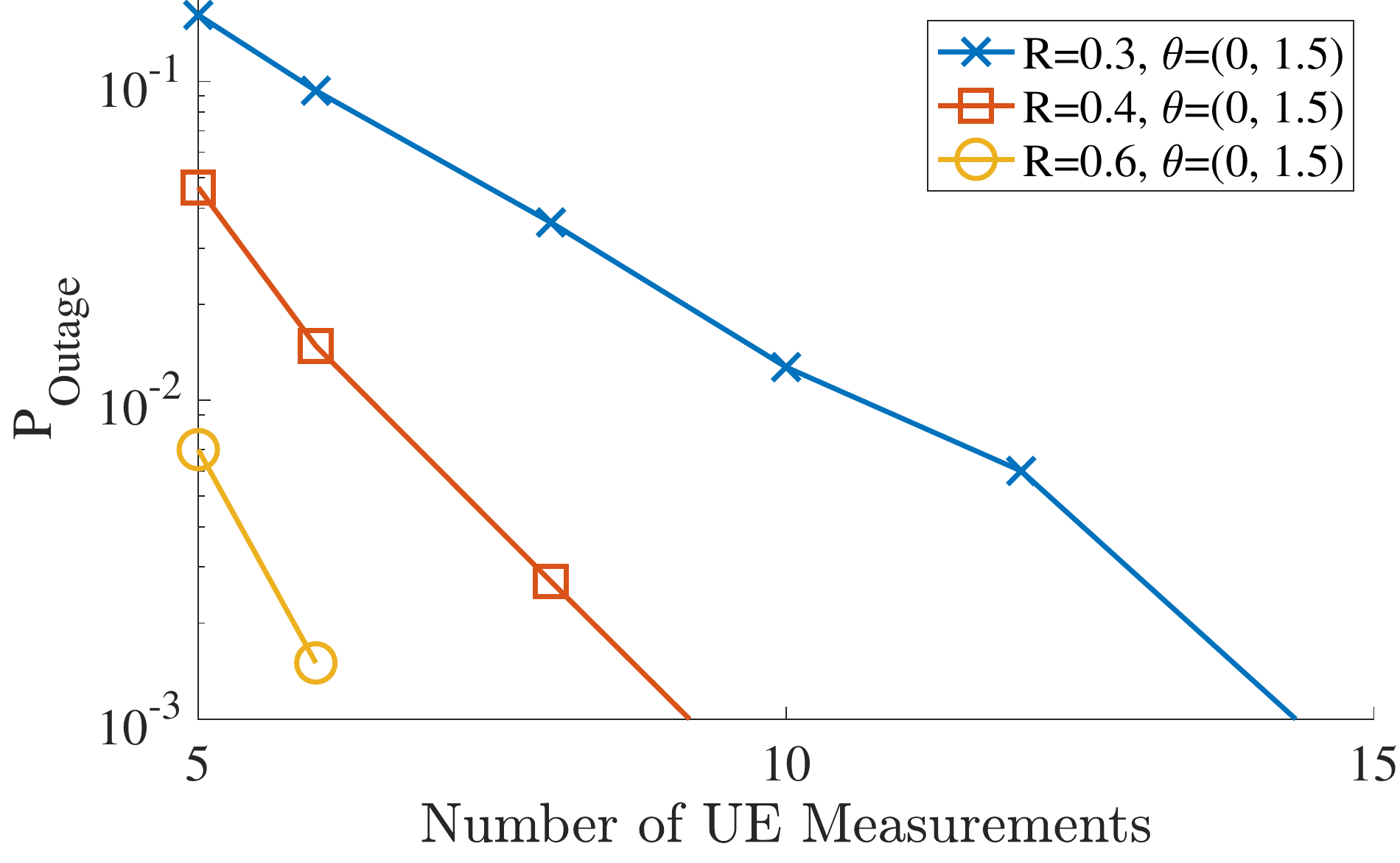}
	\caption{Outage probability of object detection algorithm.}
	\label{fig:Res2_P_O_Final} 
\end{figure}

The probability of outage of the algorithm, defined as the probability that the algorithm cannot detect the object due to a lack of blockage information, is shown in Fig. \ref{fig:Res2_P_O_Final} for different object sizes. As anticipated, this probability has an inverse relation with the object size, since larger objects have a greater chance of blocking LOS links. For example, for ten user measurements, when the object radius is $0.3$ m, the probability that the algorithm cannot find the object is around $1\%$ while this probability is lower than $0.1 \%$ for the object radius of $0.4$ m as the algorithm is provided with more B-links measurement for the same number of UE measurements.

\section{Conclusions and Future Works}
\label{Sec:Conclusion}

In this paper, a cooperative obstacle detection algorithm is proposed that exploits LOS link blockage information between user device transmitters and PDs on the ceiling to estimate the location and size of an object inside the room. We simplify and solve this problem as a quadratic linear programming problem. Simulation results show that, for specific simulation parameters, the RMS errors can be as low as  $1$ cm and $8$ cm for estimating the center and the radius of the object, respectively.

In developing the proposed obstacle detection algorithm, we assume that the blockage indicators are perfectly detected from the received signal. However, considering the noise and NLOS part of the signal, a blocked link can be detected as a non-blocked link due to strong NLOS reflections, which is considered as a missed detection. Similarly, a non-blocked weak LOS signal can be detected as a blocked link, which is considered as a false alarm for blockage detection. The proposed algorithm can be improved for practical applications to address these imperfections in the initial step of obstacle detection. Developing this algorithm for multiple objects scenario as well as addressing other realistic constraints such as limited FOV and self-blockage can also be considered as future works on this research.

\balance
\bibliographystyle{IEEEtran}
\bibliography{vlp}
\end{document}